\documentclass[titlepage,oneside,12pt]{article}

\usepackage[tiny,rm]{titlesec}
\newpagestyle{trbstyle}{
  \sethead{Sokolov, Larson, Munson, Auld, and Karbowski}{}{\thepage}
}
\usepackage{etoolbox}

\makeatletter
\patchcmd{\ttlh@hang}{\parindent\z@}{\parindent\z@\leavevmode}{}{}
\patchcmd{\ttlh@hang}{\noindent}{}{}{}
\makeatother

\titleformat{\section}{\bfseries}{\thesection. }{0pt}{\uppercase}
\titlespacing*{\section}{0pt}{12pt}{*0}
\titleformat{\subsection}{\bfseries}{}{0pt}{}
\titlespacing*{\subsection}{0pt}{12pt}{*0}
\titleformat{\subsubsection}{\itshape}{}{0pt}{}
\titlespacing*{\subsubsection}{0pt}{12pt}{*0}

\usepackage{enumitem}
\setlist[1]{labelindent=0.5in,leftmargin=*}
\setlist[2]{labelindent=0in,leftmargin=*}

\usepackage{ccaption}
\usepackage{amsmath}
\makeatletter
\renewcommand{\fnum@figure}{\textbf{FIGURE~\thefigure} }
\renewcommand{\fnum@table}{\textbf{TABLE~\thetable} }
\makeatother
\captiontitlefont{\bfseries \boldmath}
\captiondelim{\;}

\usepackage{times}

\usepackage[T1]{fontenc}
\usepackage{textcomp}

\usepackage[sort,numbers]{natbib}

\setcitestyle{round}

\usepackage{totcount}
\regtotcounter{table} 	
\regtotcounter{figure} 	

\usepackage{amsmath}
\usepackage{graphicx}
\usepackage{amsfonts}
\usepackage{float}
\usepackage{amsthm,amsmath}
\usepackage{enumerate}
\usepackage{graphicx}
\usepackage{url}
\usepackage{csvsimple}

\usepackage{xcolor}

\newcommand{\vsnote}[1]{}
\usepackage{xspace}
\newcommand{\polaris}{POLARIS\xspace}
\usepackage{tikz}
\usetikzlibrary{shapes,arrows,decorations.markings,shadows,positioning}

\usepackage[normalem]{ulem} 
 
\begin{document}
  
\title{\bf Platoon formation maximization through centralized routing and departure time coordination}
\author{Vadim Sokolov, Jeffrey Larson, Todd Munson, Josh Auld, and Dominik Karbowski\hspace{.2cm}\\
  Argonne National Laboratory\\
  \vspace*{10pt}
  4325 words + \total{figure} figures + \total{table} tables = 5825
}
\thispagestyle{empty}
\maketitle

\newpage

\thispagestyle{empty}
\section*{Abstract} 
Platooning allows vehicles to travel with small intervehicle distance in a
coordinated fashion thanks to vehicle-to-vehicle connectivity.  When applied at
a larger scale, platooning will create significant opportunities for energy
savings due to reduced aerodynamic drag, as well as increased road capacity and
congestion reduction resulting from shorter vehicle headways. However, these
potential savings are maximized if platooning-capable vehicles spend most of
their travel time within platoons. Ad hoc platoon formation may not ensure a
high rate of platoon driving.  In this paper we consider the problem of central coordination of
platooning-capable vehicles. By coordinating their routes and departure times,
we can maximize the fuel savings afforded by platooning vehicles. The resulting
problem is a combinatorial optimization problem that considers the platoon
coordination and vehicle routing problems simultaneously. We demonstrate our
methodology by evaluating the benefits of a coordinated solution and comparing
it with the uncoordinated case when platoons form only in an ad hoc manner. We
compare the coordinated and uncoordinated scenarios on a grid network with
different assumptions about demand and the time vehicles are willing to wait.

\newpage
\section{Introduction}

Platoons are composed of multiple vehicles communicating with each other and
whose movements are automatically controlled. A vehicle in a platoon knows with
good accuracy the gap with the preceding vehicle and with the leading vehicle
thanks to sensors and radio communication. Each vehicle can thus adjust its
speed with full awareness of the state of the preceding and lead vehicles,
allowing it to safely maintain a shorter gap between
vehicles~\cite{lu2011automated}. Further, researchers have shown~\cite{nowakowski2010cooperative} that drivers are comfortable with a following time-gap as low as 0.6 seconds.

Platooning also provides the prospect of reducing the massive waste incurred by
traffic congestion. A 2011 study shows urban road congestion annually costing
\$121 billion dollars based on 5.5 billion man-hours and 2.9 billion gallons of
wasted fuel~\cite{schrank2012tti}; and increasing urban population will likely
exacerbate these effects. Platooning can help alleviate congestion by addressing the highly inefficient use of
road space by human drivers: congestion occurs when vehicles occupy only 18\% of
the road for a typical highway with 2,200 vehicles per hour capacity \cite{manual2000transportation}. 
Vehicles equipped with Cooperative Adaptive Cruise Control (CACC) (which allows
for a simpler form of platooning) have been shown to improve traffic flow and
more efficiently use road
space~\cite{lu2011automated,nowakowski2010cooperative}. The impact of CACC
vehicles at different market penetration rates on a regional scale have been studied
via simulations in~\cite{vander2002effects,shladover2012impacts}, and their impact
on throughput at intersections in an urban road system was simulated in~\cite{liorisdoubling}.
The simulation studies show that CACC enables shorter following gaps and
increases road capacity from the typical 2,200 vehicles per hour to almost 4,000
vehicles per hour at 100\% market penetration. 

Platooning vehicles also use less fuel because trailing vehicles experience a
reduced aerodynamic drag. A study was conducted in \cite{tsugawa2013overview} involving three trucks driving 80 km/h with
10 m intervehicle gaps, where control
algorithms for lateral movement relied on radar measurements and
vehicle-to-vehicle communication. Analysis of their field data shows a 14\%
decrease in fuel use. Under similar speeds (60 and 80 km/h) and headway
conditions (0.3 to 0.45 seconds) a platoon of two trucks is 
studied in~\cite{bonnet2000fuel}. The trucks were connected through an electronic
system comprising a vehicle-to-vehicle controller, a tow bar controller,
and an image-processing unit. Overall, the reduction in fuel consumption ranged
from 15\% to 21\% at 80 km/h and 10\% to 17\% at 60 km/h.
In~\cite{browand2004fuel}, the authors studied fuel consumption of two 
trucks linked via an electronic control system and report 8--11\% fuel savings.
In~\cite{alam2010experimental}, the authors tested speed control algorithms for
following vehicles that use information about the road ahead sensed by the lead
vehicle. They showed a 5--8\% improvement in fuel efficiency. Computational fluid
dynamics simulations confirm field studies and show that an optimal headway
distance that minimizes drag forces is 6--8 meters; this leads to fuel savings of
7--15\%~\cite{davila2013sartre}. Studies for light-duty vehicles show
similar
savings~\cite{shida2009development,shida2010short,eben2013economy,shladover2012impacts}.

In this paper we focus on minimizing the collective fuel use of a group of
vehicles by coordinating their departure time and routing, and we then measure the fraction of total miles traveled in a platoon
for a given set of trips. An optimal routing is computed by jointly computing vehicle
routes and departure times. Routing existing platoons in a network has been
studied and solved by a number of authors using discretized optimal
control~\cite{Baskar2009a:ifac:09, Baskar2013,Baskar2009:itsc:09}, dynamic
programming~\cite{Garcia1995, Valdes2012}, and graph-based
algorithms~\cite{Doremalen2014}. These methods are applied to relatively small
networks: those with 3--10 nodes and 6--34 arcs. In contrast to our coordinated model,
the platoons in these models are not allowed to merge with other vehicles and
consequently save additional fuel; they consider only the optimal routing after
the vehicles have been grouped into platoons.

The goal of this paper is to analyze the potential improvements that can be
achieved by strategic coordinated platooning. The coordination assumes that
drivers are willing to delay their departures in order to be able to travel 
in a platoon. We analyze different levels of willingness to wait and
how such waiting affects the optimal fuel savings. We present a coordinated platooning
optimization model and evaluate the impact of optimal platoon routing by comparing
it with an ad hoc platoon formation strategy. The optimization model attempts
to minimize the collective fuel use by
routing vehicles through the network while determining when platoons should
form or dissolve. An explicit mixed-integer programming model in the GAMS
modeling language~\cite{GamsSoftware2013} and example problem data are
available at 
\begin{center}
  \url{http://www.mcs.anl.gov/~jlarson/Platooning}. 
\end{center}

The paper is organized as follows. Section~\ref{polaris} describes the
transportation system model used for opportunistic platooning simulations.
Section~\ref{model} presents our coordinated platooning optimization model.
Section~\ref{results} provides numerical results for a metropolitan road
network and compares coordinated and uncoordinated platooning with different
assumptions on travel demand and the willingness of drivers to delay their
departures. We conclude with a discussion in Section~\ref{discussion}.

\section{Opportunistic Platooning Simulations}\label{polaris}
We use \polaris, a transportation system simulator~\cite{polaris}, to simulate
ad hoc (or opportunistic) platooning. \polaris is a fully integrated, agent-based simulation
of both vehicles and traffic operations. The simulation integrates
travel demand, network simulation, and network operation models. At the center
of \polaris is a person-agent that represents travelers in the system and
their activity and travel behavior. The agents plan and schedule their
daily activities according to a variety of behavior rules and choice processes
and then travel through the network to meet their individual objectives. When
traveling from one location to another according to the behavioral objectives,
the agents choose routes through the network that minimize a personal cost
function. The agents then operate in an environment, represented by the
transportation network model, that handles movements through the system governed by
the route choice. The route can be replanned by the agent in response to network conditions,
new information, and direct system control. 

The \polaris simulator uses a variant of the Lighthill-Whitham-Richards
(LWR)~\cite{Lighthill317,richards} traffic flow
model, which is a combination of a conservation law defined via a partial
differential equation and a flow-density relation called the
fundamental diagram. The nonlinear first-order partial differential equation
describes the aggregate behavior of drivers. The model explicitly represents
the dynamics of the primary variable of interest, traffic density, which is a
macroscopic characteristic of traffic flow and the key control variable 
in transportation system management strategies. Traffic density is defined as a
number of vehicles per unit of length. The model is well studied and is
used in many transportation applications~\cite{lebacque2005first,lebacque1996godunov,hoogendoorn2001state}.

The partial differential equation underlying \polaris is solved by using Newell's
simplified kinematic waves traffic flow discretization
scheme~\cite{Newell1993281}. This is a link-based solution method and has been
recently recognized as an efficient and effective method for large-scale
networks \cite{lu2013dynamic} and dynamic traffic assignment formulations
\cite{zhang2013novel}. A notable implementation of this model is in an
open-source
dynamic traffic assignment tool DTALite~\cite{dtalite2014}. This tool is used
as the traffic simulation model agent in the \polaris framework. 

The traffic simulation model includes a set of traffic simulation agents for
intersections, links, and traffic controls.  Given a set of travelers
with route decisions and the network's traffic operation and control strategies,
the network model simulates traffic operations 
to provide capacities and driving rules on links as well as drivers' turn movements at
intersections. With these capacity and driving rule constraints, link and
intersection agents simulate the traffic flows using cumulative departures and
arrivals as decision variables based on Newell's model. This model
then determines the network performance for the route and demand models in the integrated framework. The traffic
simulation model agents also produce a set of measures of effectiveness 
such as their average speed, density, and flow rate, as well as individual vehicle
trajectories. The exact solution developed by Newell~\cite{newell} is given by
\begin{equation*}
\begin{split}
T(x,n) = \max \left( T(x_u,n) + \frac{x - x_u}{u}, \right.  
\left. ~T\left(x_d,n - \rho_{jam}(x_d - x)\right) + \frac{x_d-x}{w} \right),
\end{split}
\end{equation*}
where $T(x,n)$ is the time when vehicle $n$ crosses location $x$ on the link,
$w$ is the shock wave propagation speed, $u$ is the free-flow speed, and
$\rho_{jam}$ is the jam density of the road segment. Note that $w$, $u$, and
$\rho_{jam}$  are the parameters of the fundamental diagram. An event-based
simulation scheme is implemented by using \polaris's discrete event engine, and the
traffic flow simulator is integrated with other transportation simulation
components.

We modified the \polaris traffic flow model to account for opportunistic
platoon formation. In our study we did not simulate changes in travel demand as
a result of automation and assumed a fixed demand specified in an input trip
table. Each vehicle in the trip table is labeled as either a platoon-capable
vehicle or a regular vehicle. When simulating mixed traffic with platooning and regular vehicles, 
vehicles of both type will propagate along a link according to the LWR model. 
However, the fundamental diagram of a road link is dynamically adjusted to account for the presence of automated vehicles.  
Since the LWR model preserves the
first-in-first-out property of the traffic flow, we assume that two platoon-capable
vehicles entering the same road segment one after another will platoon on
this link. We dynamically adjust the capacity of the road segment as a function
of the number of vehicles platooning on this road segment. The capacity
adjustment factors used were derived in~\cite{vander2002effects,shladover2012impacts}. 

\section{Optimization Model}\label{model}
The set of \polaris-simulated trips is then sent to the external
optimization model in order to find optimal wait times and routes for maximizing
the time spent in a platoon.  A complete description of the optimization model
can be found in~\cite{Larson2016e}. We briefly describe the 
model variables and objective function from the optimization model. Given a collection of
vehicles and a road network described by a set of nodes and edges, our model
requires (1) the (fixed) cost to traverse any edge in the network, (2) the origin
and destination nodes for each vehicle, (3) the time each vehicle arrives in the
network, and (4) the time each vehicle must be at its destination. We assume
that the times are feasible, that is, that each vehicle's destination time
is at least their origin time plus the shortest path time from its origin to
its destination. For our simulations, we assume vehicles are willing to wait a
short period of time at their origin nodes provided they can save fuel by
platooning, but we do not allow vehicles to wait at intermediate nodes.

Given a problem instance defined by these parameters, the
optimization model chooses routes and departure times for each vehicle so
that the collective fuel use is minimized while ensuring that each vehicle reaches its
destination on time. If $n$ vehicles travel on the same road segment at the
same time, $n-1$ use 10\% less fuel than the remaining vehicle (which is
assumed to be leading the platoon). 

Our objective is to minimize the
overall fuel consumed. We use a simple assumption that the
amount of fuel consumed by a vehicle while traversing an edge $(i,j)$ is
constant, and we denote it by $C_{i,j}$. We denote the delay in departure time of vehicle
$v$ at its origin by $t_v$, the fraction of fuel saved by platooning
by $\eta$, and the cost of waiting by each vehicle by $\epsilon_v$. For our
study, $C_{i,j} = 1$ for all edges in the grid and $\eta = 0.1$. If the
decision variable $f_{v,i,j}=1$ when vehicle $v$ takes $(i,j)$ and
$q_{v,w,i,j}=1$ if vehicle $v$ follows vehicle $w$ on $(i,j)$, then the
objective function is
\begin{equation}\label{eq:objective}
\sum_{v,i,j} C_{i,j} \left( f_{v,i,j} - \eta \sum_{w} q_{v,w,i,j} \right) +
\epsilon_v t_{v}.
\end{equation}
In our current study we focus on the maximal possible savings and set $\epsilon_v = 0$. 
For fleet managers coordinating the routes of many vehicles,
$\epsilon_v$ should be the cost per unit time for a stationary vehicle: the
drivers' wages plus any idling costs. Such a straightforward calculation is less
obvious for private drivers. Also, individuals may need additional incentives
in order to be willing to add even a short period of time to their commutes in
order to reduce their fuel use by 10\%.

Note that a naive implementation of the prescribed model will quickly
become computationally intractable because of, for example, generating binary
variables $q$ for all pairs of vehicles and all edges in the network. A more
systematic approach, used in the available code and discussed in depth
in~\cite{Larson2016e}, is to generate variables only when necessary. A
vehicle will not travel more that $\frac{1}{1-\eta}$ times its shortest-path route
between its origin and destination~\cite[Lemma 2.2]{Larson2016e}; therefore,
most $f_{v,i,j}$ can be removed for most edges in a real-world network.
Similarly, $q_{v,w,i,j}$ need exist only if vehicles $v$ and $w$ can possibly
traverse edge $(i,j)$ simultaneously given their origin/destination times. Such
considerations dramatically reduce the model size.

Naturally, this model requires a collection of constraints. Any vehicle that
enters a nondestination node must exit it.  If vehicles are platooning on an
edge, the times they enter the edge must be equal. A vehicle cannot enter
another edge until it has traversed its current edge. For a thorough discussion
of the model and constraints, see~\cite{Larson2016e}.

\section{Case Study}\label{results}
\begin{figure}[H]
  \centering
  \includegraphics[width=0.45\linewidth]{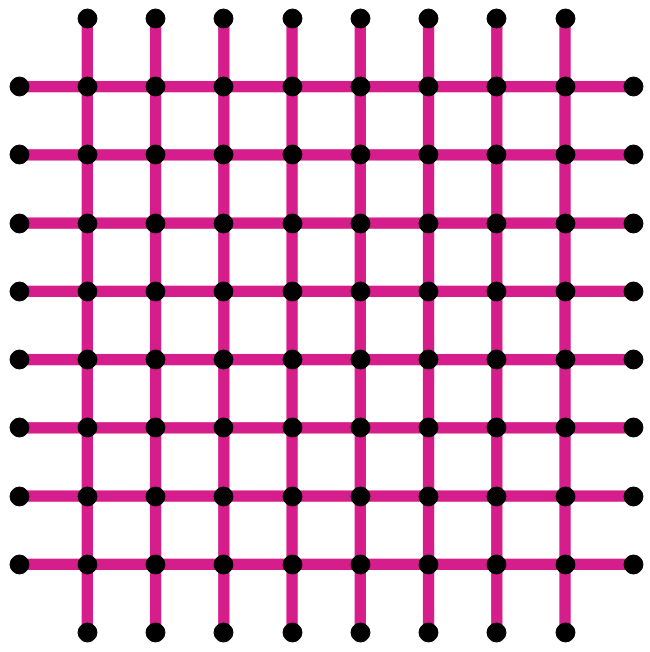}
  \caption{Network under consideration: a $10\times10$ grid \label{fig:networks}}
\end{figure}

To test the effects of coordinated and uncoordinated platooning, we
performed experiments on the $10\times 10$ grid shown in Figure~\ref{fig:networks}, in which each link has a length of
1 km. Even though the grid model network used in this study appears simple, finding the optimal
solution on such grid network is more challenging than on a real highway network since many different routes of
the same length exist between most pairs of origin/destination nodes. The number of
shortest paths between $(0,0)$ and $(m,n)$ in a grid is ${m+n \choose n}$, whereas on a
highway network there are usually a very small number of valid alternative
routes between an origin and a destination (usually not more than two). In this
case study we assume no congestion on the network, and the cost of traversing a road
link is assumed to be proportional to free-flow travel time on this link.

\begin{figure}[H]
  \centering
  \includegraphics[width=0.7\linewidth]{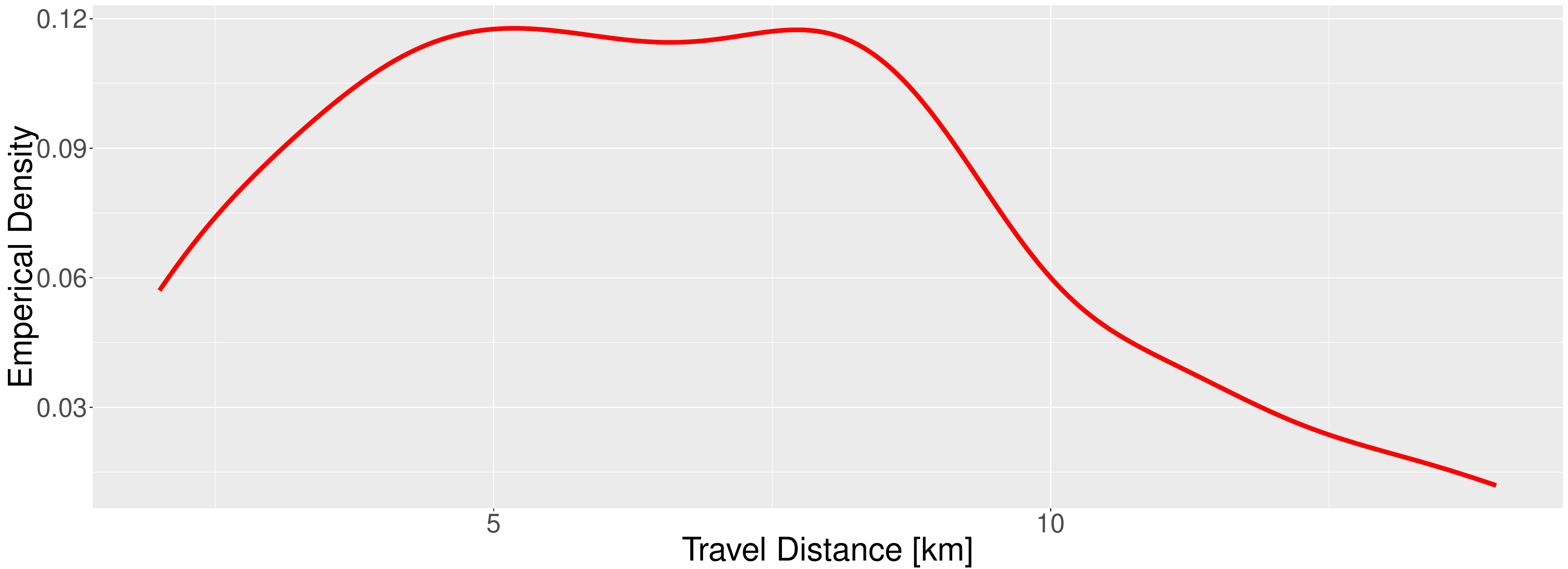}
  \caption{Distribution of trip length}
  \label{fig:vkt-density}
\end{figure}

Origins and destinations are randomly generated for 50 vehicles.
The trip length distribution is shown in Figure~\ref{fig:vkt-density}; its mean is 7 km.
We make the simplifying assumption that all 50 vehicles can 
be rerouted and controlled and that their coordination does not affect link travel
times. This will not hold as more vehicles are routed, but it is a valid
assumption when small percentages of vehicles are under control.

\begin{figure}[H]
  \begin{tabular}{ccc}
    \includegraphics[width=0.3\linewidth]{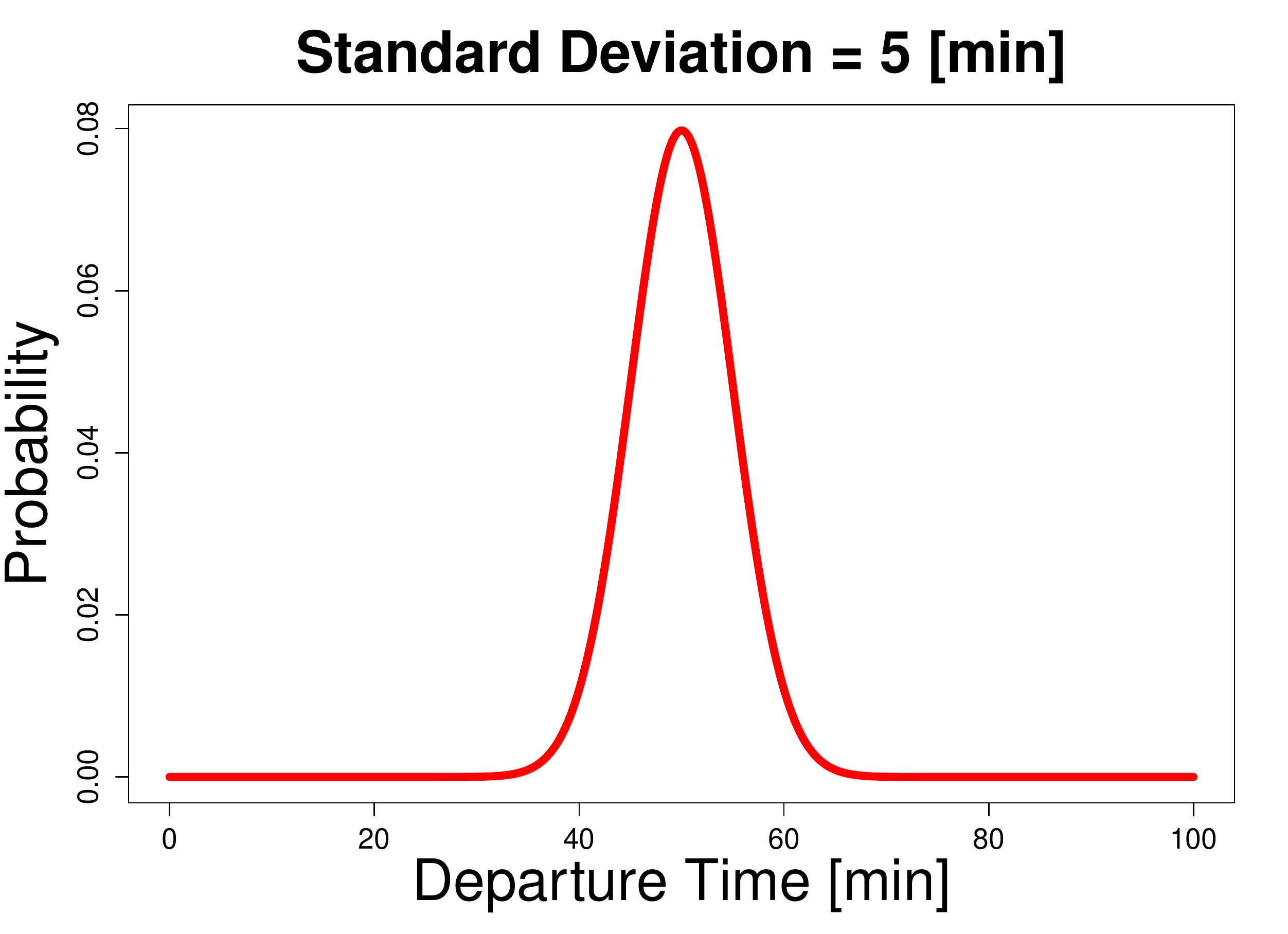} & \includegraphics[width=0.3\linewidth]{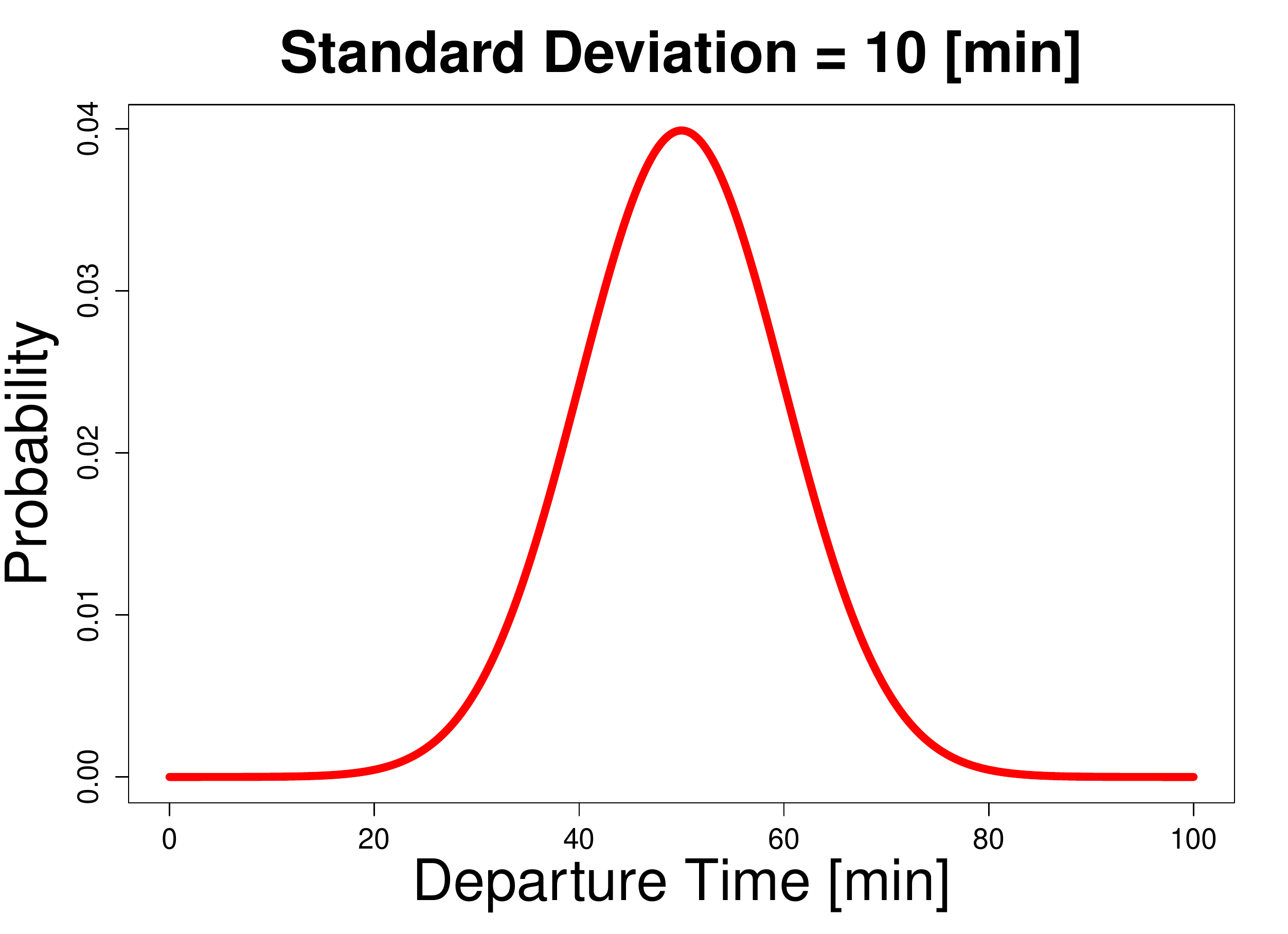} & \includegraphics[width=0.3\linewidth]{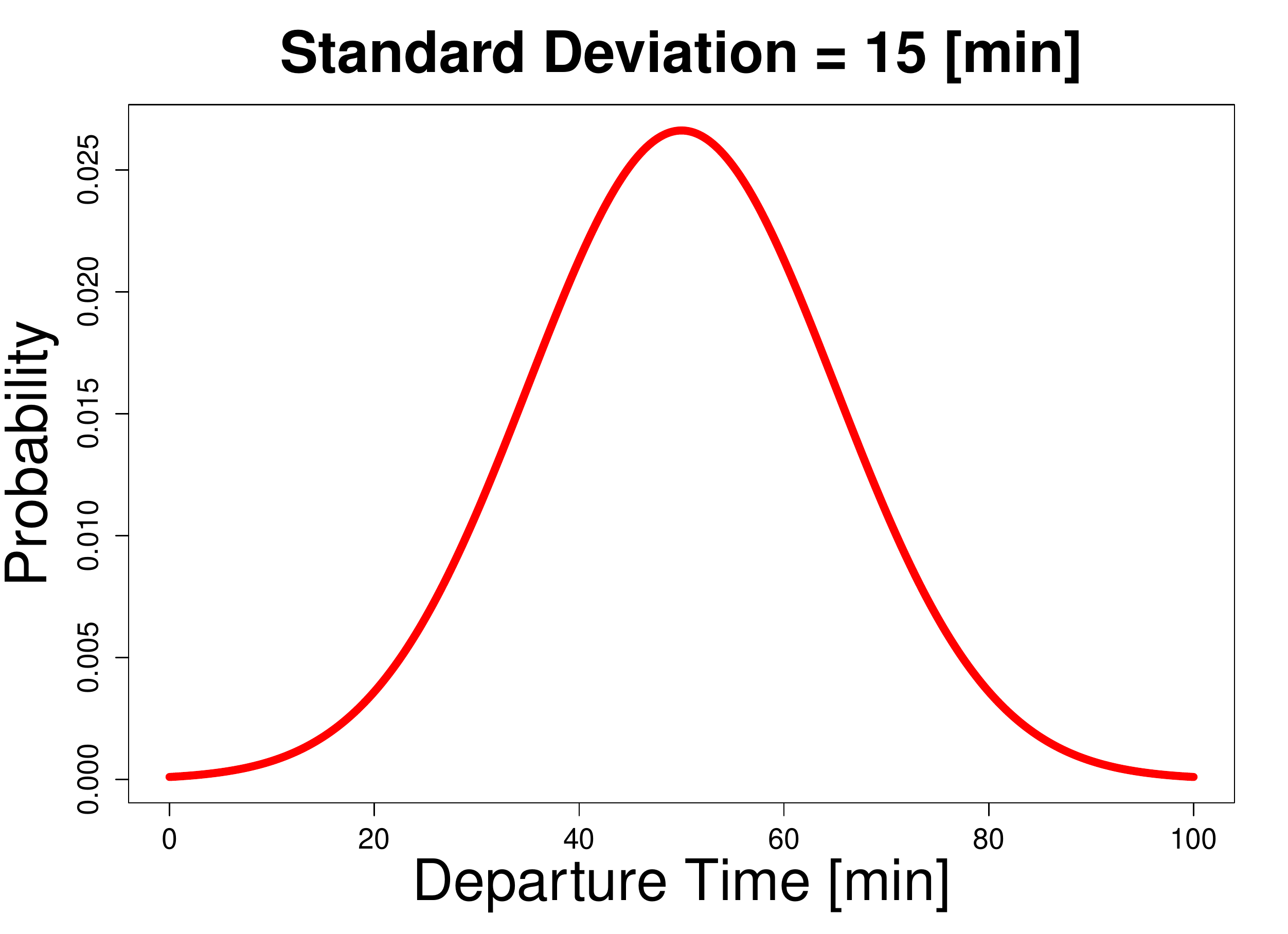} \\ 
    \includegraphics[width=0.3\linewidth]{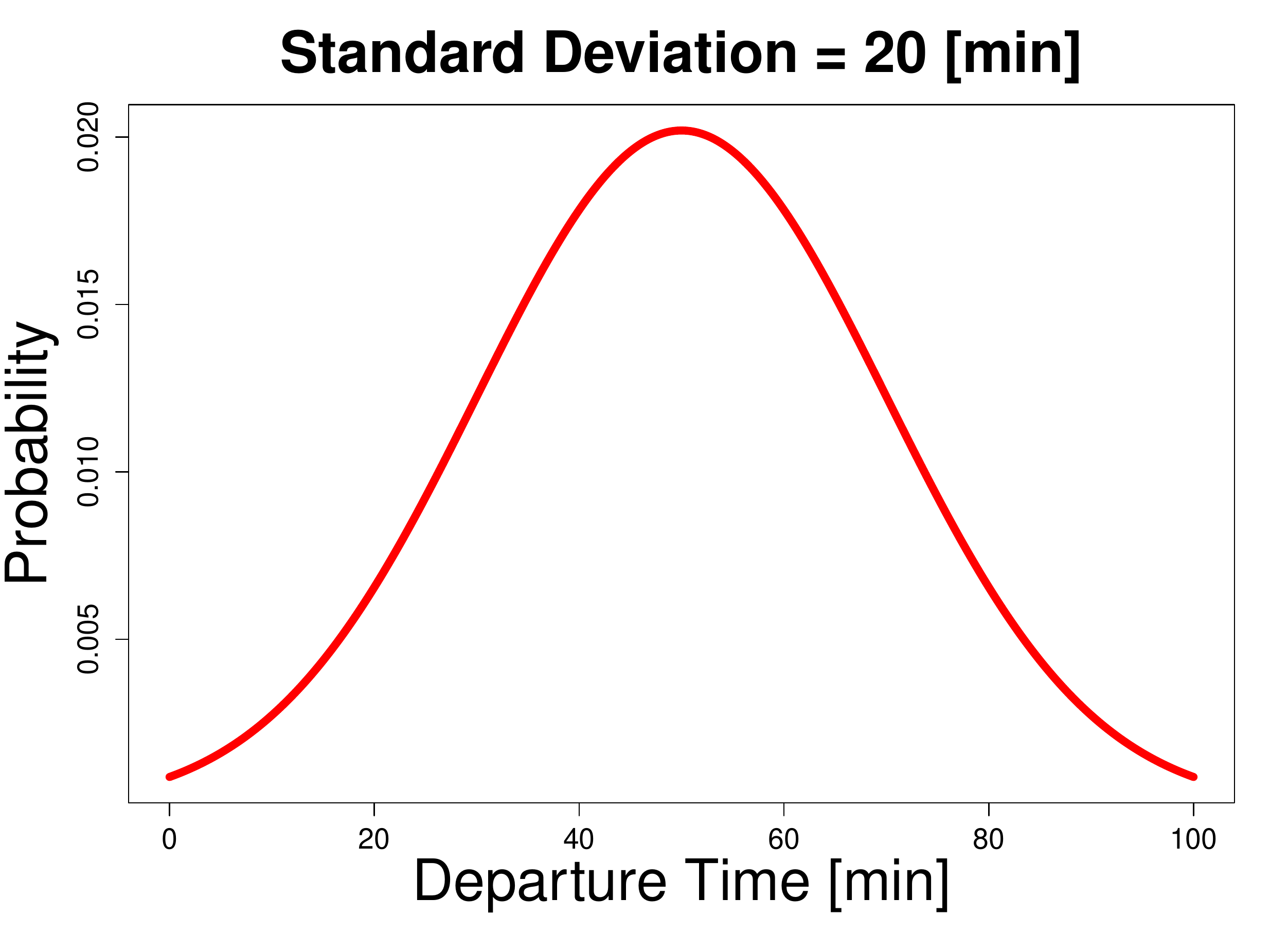}& \includegraphics[width=0.3\linewidth]{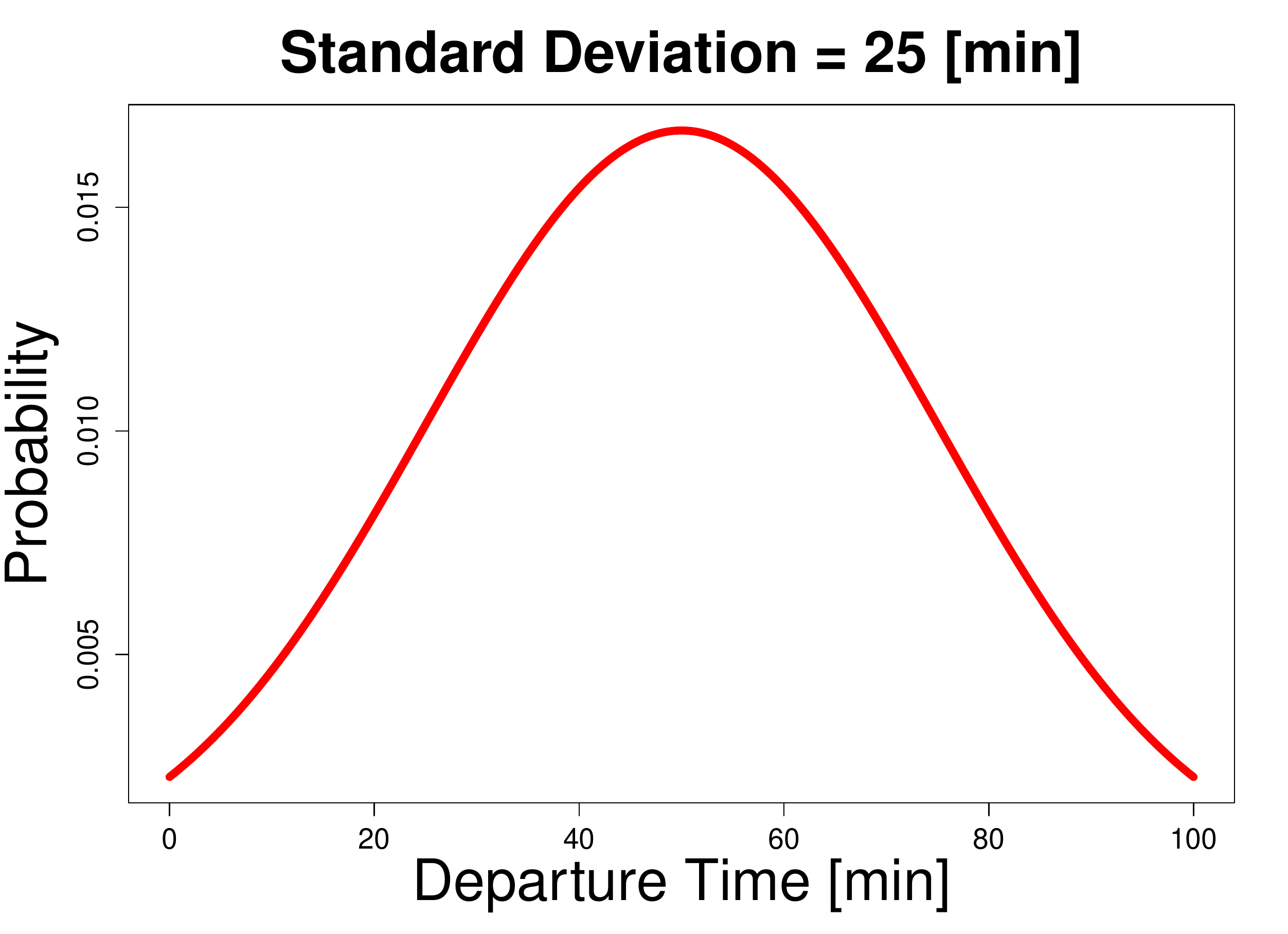} & \includegraphics[width=0.3\linewidth]{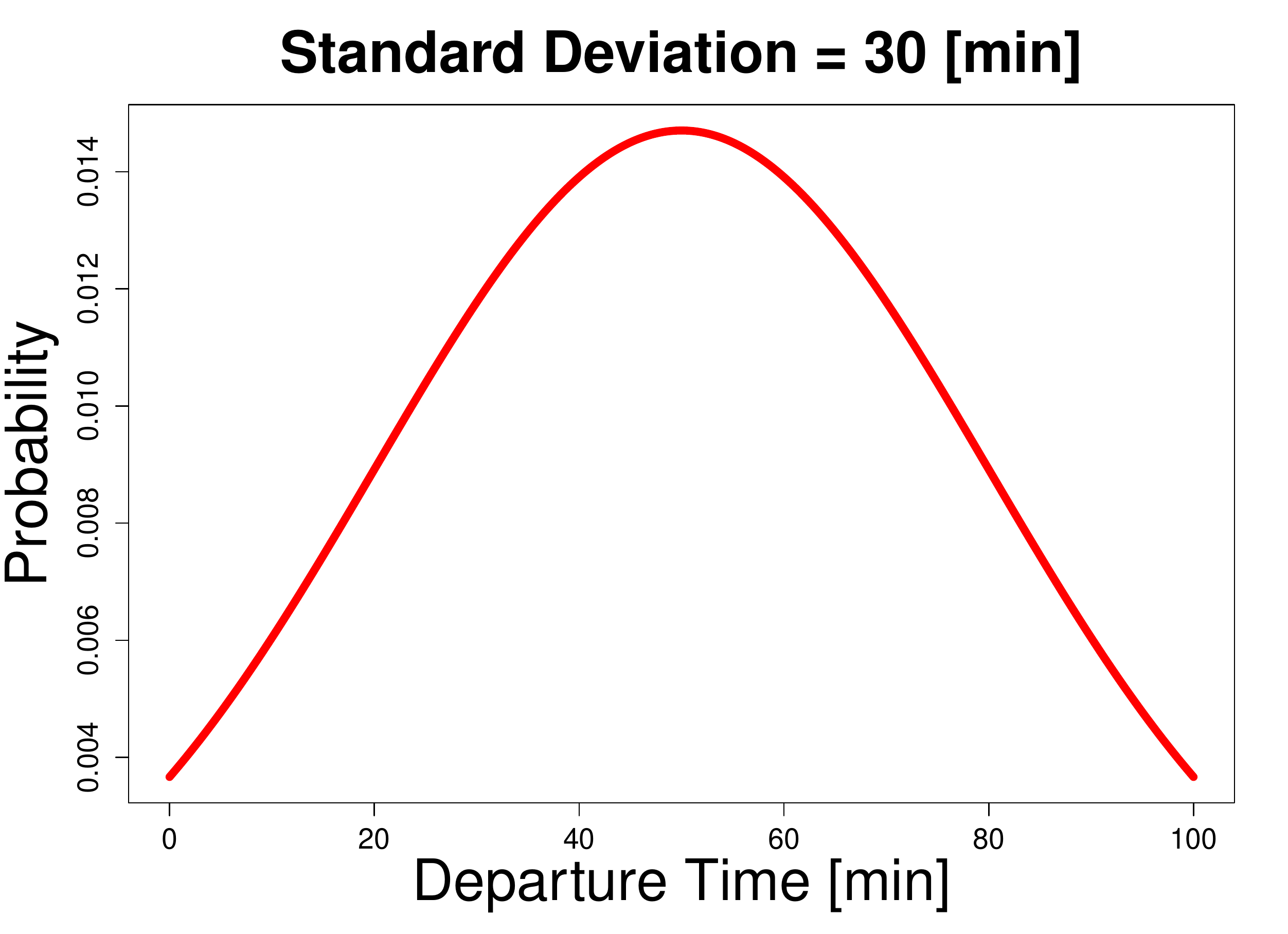} \\ 
  \end{tabular} 
  \caption{Six scenarios for departure time distribution}
  \label{fig:to-density}
\end{figure}

The departure times $T^O_v$ for each vehicle $v$ are randomly drawn
from a truncated normal distribution with support of $[0,100]$, mean 50, and 6
different standard deviations.  The departure time distributions are 
shown in Figure~\ref{fig:to-density}. Each vehicle must arrive at
its destination at time $T^D_v$, set to
\begin{equation}\label{eq:dest_time}
  T^v_D = T^v_O + T_{O_{v},D_{v}} + p,
\end{equation}
where $T_{O_v,D_v}$ is the minimum time between the vehicle's origin and
destination and $p$ is some pause time. We assume that trailing vehicles in a
platoon use 10\% less fuel than do vehicles leading a platoon or traveling
alone on a given edge. 

One of the most important parameters for maximizing driving in platoons
is $p$, the upper bound on the amount of time vehicles are
willing to wait. Naturally, the longer vehicles are willing to wait, the more
platooning possibilities exist, and therefore platooning can occur. If the pause time
$p$ in \eqref{eq:dest_time} is zero, then every vehicle must travel from its
origin to its destination along its shortest path and can participate only in ad hoc
platoons. This scenario corresponds to the uncoordinated platooning case. If $p > 0$, a
vehicle can wait to lead/follow another vehicle, thereby decreasing the
collective fuel use. In our experiments, increasing $p$ past a certain value
provides no additional savings. We simulated our case study with five 
maximum possible wait times: $p \in \{0,10,20,30,40\}$. We ran Gurobi for five
minutes on each GAMS model of each problem instance. 

For our case study we use wait time and the vehicle-miles-traveled (VMT) ratio to estimate the efficiency
of the optimal routing when compared with opportunistic platooning. The VMT ratio
is the ratio of miles driven in a platoon to the total miles driven by a
vehicle. 
\[
\mathrm{VMT~ratio} = \dfrac{\mathrm{Platoon~VMT}}{\mathrm{Total~VMT}}
\]

\begin{figure}[H]
\begin{tabular}{cc}
	\includegraphics[width=0.5\textwidth]{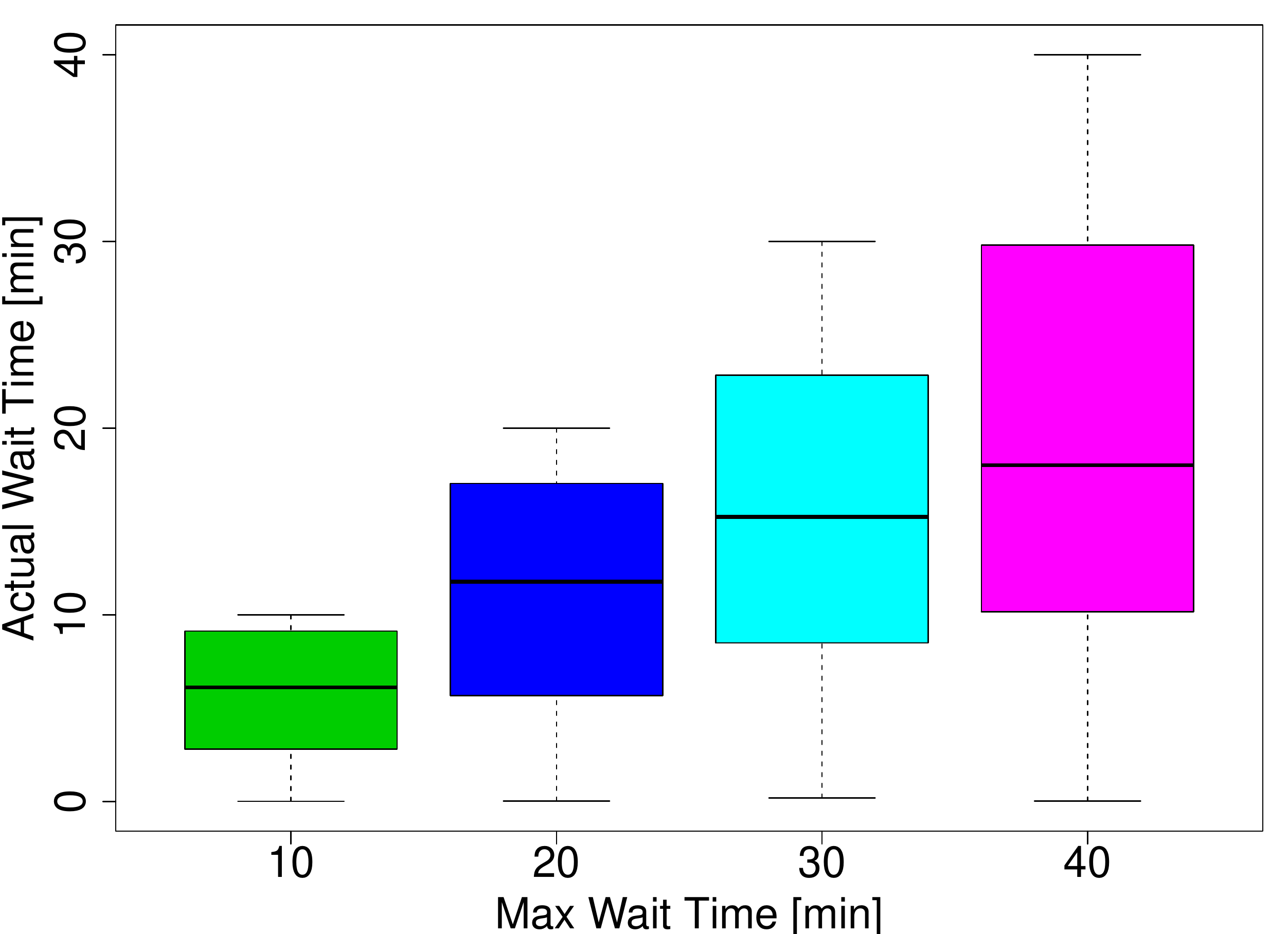}&
	\includegraphics[width=0.5\textwidth]{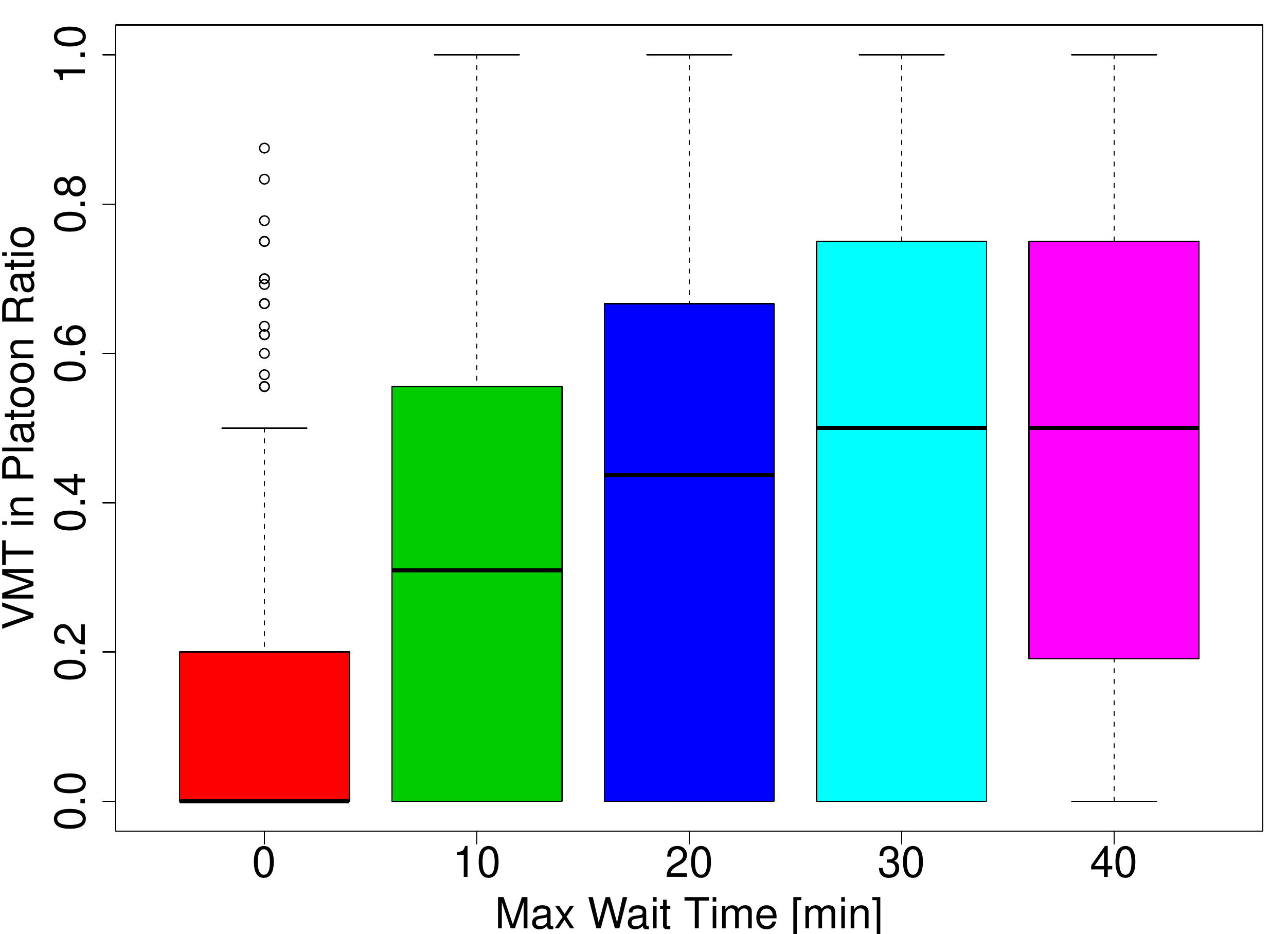}\\
	(a) Actual wait times & (b) Ratio of distance driven in platoon
\end{tabular}
\caption{Box plots comparing the impact of coordinated platooning for
different assumptions on maximum allowed wait time. Panel (a) compares
actual wait times  experienced by drivers. Panel (b) compares the ratio
of distance driven in a platoon.}
\label{fig:box_plots}
\end{figure}


Figure~\ref{fig:box_plots} shows a summary of the simulated results for the case
study.  We compare results for different assumptions about the maximum wait time. This 
parameter controls the amount of time by which each driver is
willing to delay his/her departure time, with zero corresponding to the
opportunistic platooning scenario; that is, drivers depart at the originally
intended time and platoon only in an ad hoc fashion. For each
scenario we calculate two metrics: the ratio of distance driven in a
platoon and the average wait time. Naturally, the average wait time is less than
maximum wait time and is zero for opportunistic platooning scenario, so it is
not shown on Figure~\ref{fig:box_plots}(a). 

The average platoon distance ratio for opportunistic platooning is 0.12. On
the other hand, for coordinated platooning when we set the maximum wait
time to 10, the average distance-in-platoon ratio is nearly tripled to 0.32.
Note that the average wait time
for this scenario is 5 minutes, which is well below the upper bound of 10
minutes.

The largest gain in the platoon distance ratio is when we switch from the opportunistic platooning
scenario (maximum wait time = 0) to the optimal platooning with a maximum wait
time of 10 minutes. For a maximum wait time larger than 10 minutes we do not see
significant improvement in the ratio. 

\begin{figure}[H]
  \begin{tabular}{ccc}
  	\includegraphics[width=0.5\linewidth]{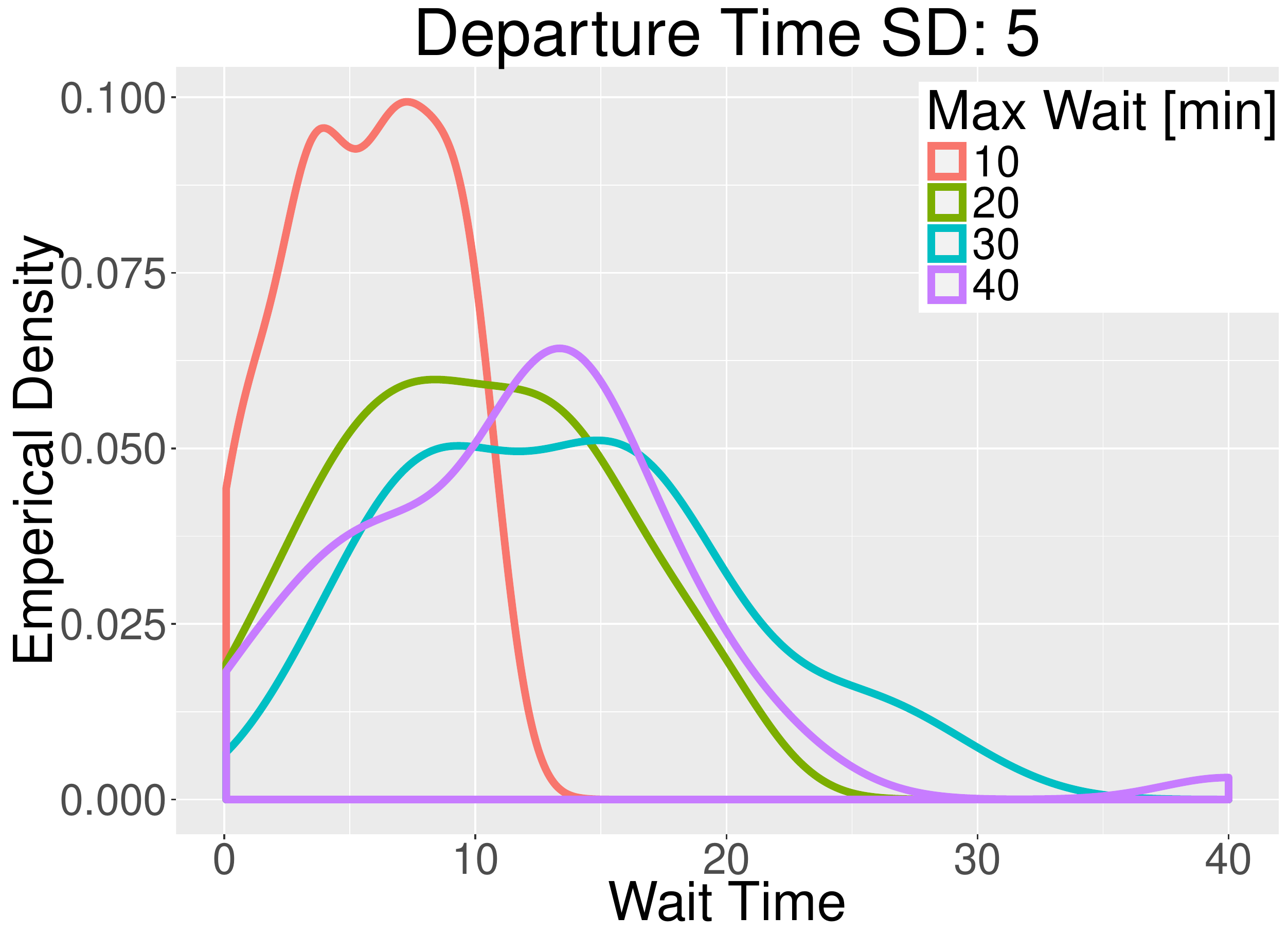} &
  	\includegraphics[width=0.5\linewidth]{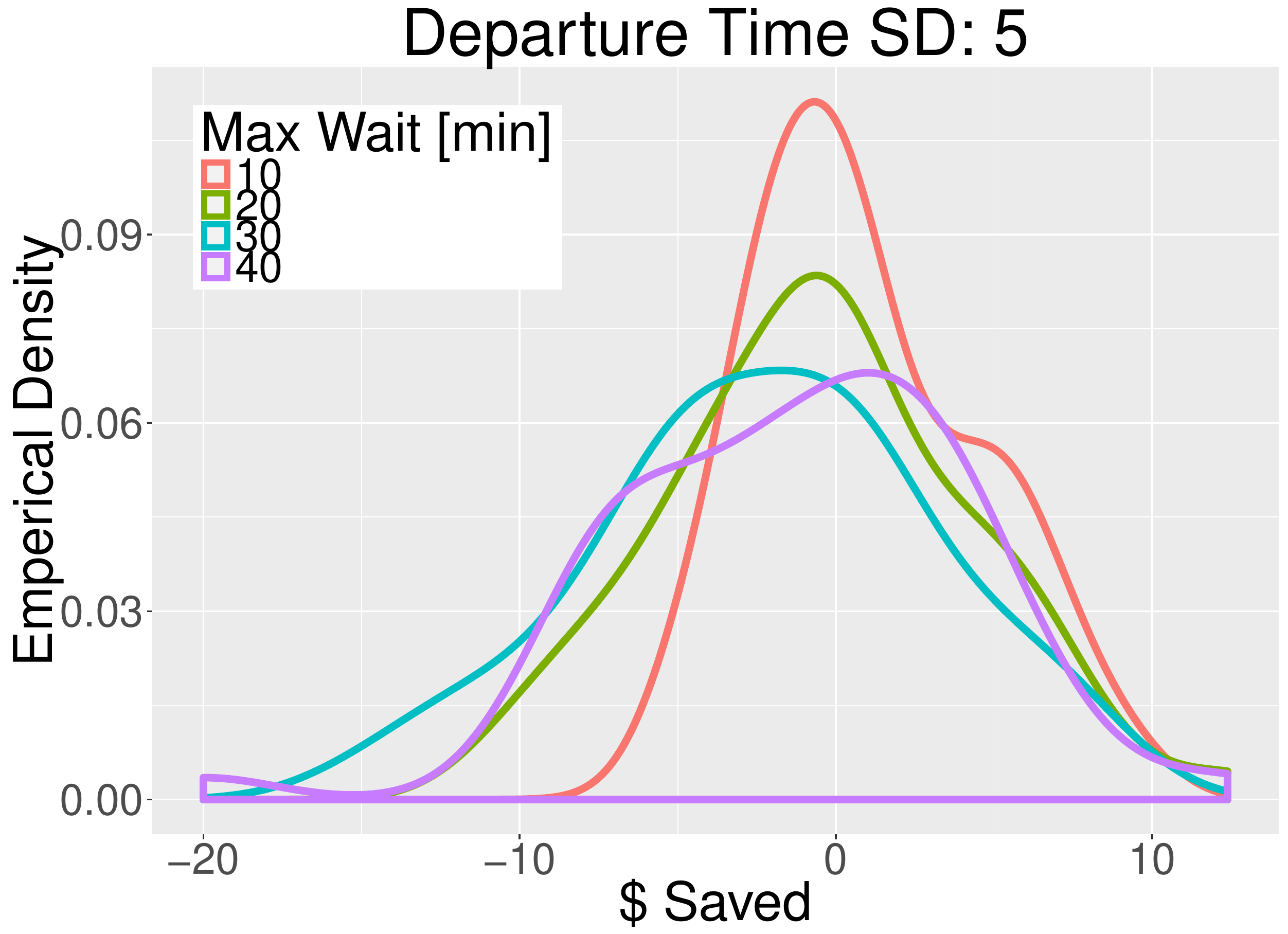} \\
  	(a) Wait Time Distribution & (b) Dollar Savings 
  \end{tabular} 
  \caption{Distribution of departure time delays calculated by optimization algorithm and economic savings for the system users}
  \label{fig:savings-density}
\end{figure}

However, the benefits of platooning  (i.e., energy savings) must be traded with the
extra wait time required under strategic routing of automated
vehicles scenario. Making assumptions about the mean fuel consumption (gallon/miles)
and value of time (\$/hour), one can calculate savings associated with the
strategic routing by 
\[
\mbox{Savings} = \mbox{VMT in Platoon}\times \eta \times \mbox{Fuel Consumption} \times \mbox{Fuel Cost}-  \mbox{Wait}\times \mbox{Value of Time}.
\]
Here $\eta$ is the ratio of fuel saved while driving in a platoon. Assuming $\eta 
= 0.1$, $\mbox{Fuel Consumption} = 0.04$ gallons/miles, (equivalent to 25 mpg),
$\mbox{Value of Time} = 30$ \$/hour, and $\mbox{Fuel Cost} = 3$ \$/gallon we
calculate the distribution of the savings associated with the
centralized routing strategy. Figure~\ref{fig:savings-density} shows the
results for different assumptions about maximum wait time
and departure time distributions. 

We can see from Figure~\ref{fig:savings-density}(a) that the wait time
distribution does not change for scenarios with wait times greater than 10,
hence there exists some threshold beyond which an increase in the wait time
does not
bring benefits. On the other hand, when we contrast the wait time with the net
economic benefit shown in Figure~\ref{fig:savings-density}(b), we see that
under our assumptions, it is negative for all of the users. Thus, for such a
system to be viable, additional benefits should be associated with
centralized routing strategy. Examples of such benefits might include
saved travel times as a result of reduced congestion or incentives for CACC
drivers such as reduced tolls or access to dedicated lanes. Assessing the impact of
centralized routing strategies on the systemwide congestion levels is the
direction of our future research.

\begin{figure}[H]
  \centering
  \begin{tabular}{cc}
    \includegraphics[width=0.45\linewidth]{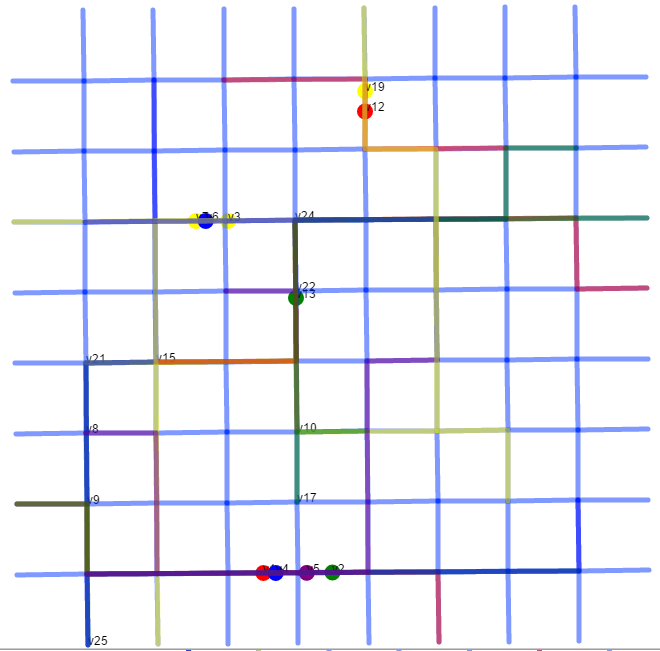} & \includegraphics[width=0.45\linewidth]{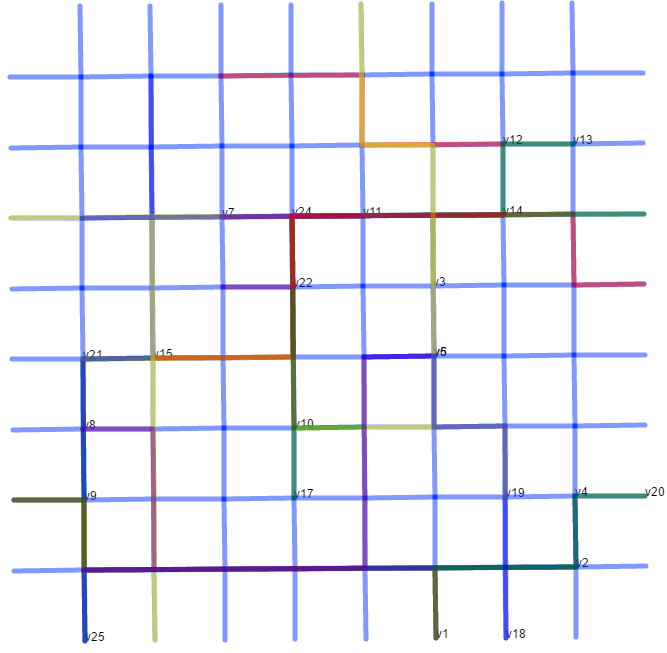} \\ 
  \end{tabular} 
  \caption{Snapshot for optimal platoons}
  \label{fig:animation}
\end{figure}

As part of the analysis framework we developed a web-based animation of the
optimization results for the grid network. Snapshots of the animated
visualization are shown on Figure~\ref{fig:animation}.
Animation of the solutions on an instance of the optimization problem with 25 vehicles is available at \url{http://polaris.es.anl.gov/cav_map_notiles/}.


\section{Discussion}\label{discussion}
In this paper we demonstrated a method of coordinating platoon formation in
which vehicles routes and departure times are strategically chosen by a
centralized authority. 
 We showed that at reasonable waiting times, one can
substantially increase the distances vehicles travel in a
platoon when a coordinated approach is used. To our knowledge this is one of
the first papers that presents a relatively large-scale case study of
coordinated platooning and compares such an approach with the uncoordinated case. We used
Gurobi to solve GAMS models to find optimal routes and
departure times. We used the transportation system modeling framework
\polaris to simulate the uncoordinated platooning case. \polaris can  
simulate large-scale transportation systems with millions of trips
in a matter of hours. Thus, it can be used to analyze the impacts of opportunistic
platooning for large-scale models. However, the underlying
optimization problem for coordinated platooning is the combinatorial optimization
that currently scales poorly with the number of vehicles. Certain assumptions and modeling tricks have 
allowed us to solve problem instances with 50 vehicles for a fairly complicated network.

Current research includes considering possible heuristic rules in order 
to improve the solution times of the optimization model on larger problems, with the goal of
solving instances with thousands of vehicles. \polaris is also being adjusted to
model a mix of platooning and nonplatooning vehicles. Provided that vehicles are 
traveling at free-flow speeds, the simulation setup and optimization model
are still accurate. The congested network case is considerably more difficult
to address. We are exploring using the optimization model as an open-loop
controller to feed into \polaris. When the network is congested, care is being
taken to ensure that the routes produced by the optimization model are feasible and
converge to a stable routing. We are also working to relax the
optimization model assumption
that platooning vehicles travel at free-flow speeds in order to accurately
analyze congested networks. Moreover, we are
analyzing fuel savings using the high-fidelity vehicle energy model
AUTONOMIE~\cite{aut,auld2016disaggregate} in order to better understand the
impacts of coordinated platooning. Considering the impact of traffic lights on
platoon formation and energy savings~\cite{liorisdoubling} is another direction
for future research.

\section*{Acknowledgements} 

We are grateful to comments from four anonymous reviewers that greatly improved
an early version of this manuscript.  This material is based upon work
supported by Laboratory Directed Research and Development (LDRD) funding from
Argonne National Laboratory, provided by the Director, Office of Science, of
the U.S. Department of Energy under contract DE-AC02-06CH11357.

\bibliographystyle{trb}
\bibliography{../../../bibs/platooning}

\clearpage 

\framebox{\parbox{.92\linewidth}{The submitted manuscript has been created by
UChicago Argonne, LLC, Operator of Argonne National Laboratory (``Argonne'').
Argonne, a U.S.\ Department of Energy Office of Science laboratory, is operated
under Contract No.\ DE-AC02-06CH11357. The U.S.\ Government retains for itself,
and others acting on its behalf, a paid-up, nonexclusive, irrevocable worldwide
license in said article to reproduce, prepare derivative works, distribute
copies to the public, and perform publicly and display publicly, by or on
behalf of the Government.}}
\end{document}